# Excellent catalytic performance towards the hydrogen evolution reaction in topological semimetals

Lirong Wang[a,b,1], Ying Yang[c,1], Jianhua Wang[d,1], Wei Liu[a], Ying Liu[a,b], Jialin Gong[d], Guodong Liu[a,b],

Xiaotian Wang[d,*], Zhenxiang Cheng[e,*], Xiaoming Zhang[a,b,*]

[a]*State Key Laboratory of Reliability and Intelligence of Electrical Equipment, Hebei University of Technology, Tianjin 300130, China*
[b]*School of Materials Science and Engineering, Hebei University of Technology, Tianjin 300130, China*
[c]*College of Physics and Electronic Engineering, Chongqing Normal University, Chongqing, 401331, China*
[d]*School of Physical Science and Technology, Southwest University, Chongqing 400715, China*
[e]*Institute for Superconducting and Electronic Materials (ISEM), University of Wollongong, Wollongong 2500, Australia*
[1]*These authors contributed equally to this manuscript.*
E-mail: *xiaotianwang@swu.edu.cn, cheng@uow.edu.au, zhangxiaoming87@hebut.edu.cn*

Topological states of matter and their corresponding properties are classical research topics in condensed matter physics. Quite recently, the application of materials that feature these states has been extended to the field of electrochemical catalysis and become an emerging research topic that is receiving increasing interest. In particular, several recent experimental studies performed on topological semimetals have already revealed high catalytic performance towards hydrogen evolution reaction (HER), strongly promoting acceptance of the fresh concept of the topological catalyst. This emerging concept has experienced rapid developments in the last few years, but these developments have been rarely summarized. Herein, we offer a comprehensive review on the state-of-the-art progress in developing topological catalysts for the HER process through topological semimetals such as Weyl semimetals, Dirac semimetals, nodal line semimetals, etc. The course of development, the general research routes, and the fundamental mechanisms in topological catalysts are also systematically analyzed in this review.

## 1. Introduction

Over the past few decades, the world has been experiencing a significant increase in energy consumption due to major technological advances paired with a population boom.[1] Currently, the primary energy source worldwide remains fossil fuels. The consumption of fossil fuels has caused serious environmental issues.[2] Against this backdrop, increasing efforts have been made towards developing alternative energy sources that are cleaner and more sustainable than fossil fuels.[3] Hydrogen is one such clean energy resource, because it can be produced by water splitting with a small amount of external energy, and the main products would turn back to water after combustion.[4] To date, water splitting has been realized by different approaches, such as electrolysis, thermolysis, photocatalysis, etc.[5-13] Electrolysis in particular has had the most success, benefiting from the great progress in developing high performance electrocatalysts that has been realized in recent years.[14-17] Commercial-scale production of hydrogen imposes high requirements on electrocatalysts in terms of their catalytic activity, stability, and cost.

In the electrocatalytic process of water splitting, the reaction that takes place at the anode is known as the hydrogen evolution reaction (HER).[18] Noble metals (and noble-metal-containing materials), such as Pt, Ir, and Pd show high electrocatalytic performance towards the HER, but their scarcity and corresponding high costs greatly limit their commercial applications.[19-21] Therefore, intense research attention has been paid to exploring low-cost electrocatalysts based on earth-abundant materials such as transition metal chalcogenides,[22,23] transition metal phosphides,[24,25] transition metal carbides,[26,27] transition metal nitrides,[28,29] etc. Although their cost has gone sharply down, most catalysts based on them are still far from satisfactory due to their much lower HER activity and worse catalytic stability than noble metals. Different treatments have emerged in response and been employed to increase the catalytic activity and stability of catalysts, including strain engineering,[30] phase engineering,[31] defect engineering,[32] the effects of electric field, magnetic field, and light,[33] heterostructure formation,[34] heteroatom doping,[35] etc.

The discovery of topological materials has provided more possibility to develop high performance electrocatalysts.[36,37] Among the various topological concepts, the topological insulator, which is insulating in the bulk but carries robust conducting states on the edge or surface, is a typical category with attractive quantum effects in solid-state physics.[38-40] These novel characteristics endow topological insulators with great potential for various applications in quantum computing, electronic storage, and topological refrigeration.[41-43] In addition, topological insulators were also demonstrated as suitable for providing excellent support for heterogeneous catalysis.[44-46] In the last few years, with the research interest on topological aspects extended to topological semimetals,[47-52] the concept of the topological catalyst has been greatly promoted. Unlike topological insulators, the bulk states of topological semimetals are conductive to ultrahigh electronic mobility, which is believed to favor the catalytic process.[36,37] In addition, topological semimetals are found to carry a large number of topological emergent particles including different categories of nodal points, nodal lines, and nodal surfaces.[53-55] These topological emergent particles also show diverse surface states. The main concept of topological semimetals as high performance catalysts was motivated by their outstanding chracteristics, combining the topological bulk states and nontrivial surface features, which include excellent electron conductivity, highly-active surface states, extrahigh electron mobility, and proper adsorption energy for molecules, as evidenced by both theoretical and experimental studies.[56-67]

Herein, the application of various topological semimetals to high performance catalysts for the electrochemical HER process is systematically summarized in this review. First, the backgound for the fundamental physics of topological states of matter is expounded. The differences of diverse topological fermions and their corresponding topological surfaces are discussed. Then, the feasiblity of topological semimetals as HER catalysts is fully demonstrated from the experimental breakthrough in Weyl semimetals. The design of hybrid Weyl semimetals with a long Fermi arc for further catalytic enhancement is also discussed. In addition, the realizations of other nodal point semimetals, including Dirac and high-charge chiral topological semimetals, as HER catalysts are further analyzed. Besides these nodal point topological catalysts, a discussion of the recent progress in nodal line topological catalysts is also provided. Finally, a general summary for the research actualization of the topological catalysts is provided to give researchers an overall understanding of this emerging research field.

## 2. Background

### 2.1 Basic concept of topological semimetals

We start with the introduction of the Weyl semimetal. This refers to a material that hosts Weyl points around the Fermi level,[68] as shown in Figure 1A. Akin to the Weyl fermions in high-energy physics, a Weyl point is formed by the linear crossing of two bands. Hence, a Weyl point has twofold degeneracy, which could be generally described by:

$$H_{WSM}(\boldsymbol{k}) = \boldsymbol{w} \cdot \boldsymbol{k}\,\sigma_0 + v_x k_x \sigma_x + v_y k_y \sigma_y + v_z k_z \sigma_z \qquad (1)$$

Where the wavevector $\boldsymbol{k}$ is measured from the crossing point. $\sigma_0$ is a $2 \times 2$ identity matrix, and the $\sigma's$ refer to the Pauli matrix. The first term in Eq. (1) stands for the overall shift of the Weyl point. In particular, this term could tilt the Weyl cone, resulting in a type-II Weyl point if $|w_i| > |v_i|$ (i=x, y, z).[69] Otherwise, the Hamiltonian (1) describes a type-I Weyl point with an untilted Weyl cone. In realistic materials, there still has a possibility with coexisting both type-I and type-II Weyl points in a single material, which is known as a hybrid Weyl case.[70,71] It should also note that a three-dimensional (3D) Weyl point is a robust band crossing, and besides the translation symmetry of crystals, it does not require any extra symmetry.

From the topological perspective, a Weyl point can be regarded as a source or drain for the Berry curvature field. The Berry curvature[72] is defined based a gauge potential Berry connection, $A(\boldsymbol{k}) = i\langle u_n | \nabla_k u_n \rangle$,[73] and for concreteness, $\Omega_n(\boldsymbol{k}) = \nabla_{\boldsymbol{k}} \times A(\boldsymbol{k})$. Thus, by integrating the Berry curvature field on a closed surface that encircles the Weyl point, one could derive a net topological charge,

$$N_S = \frac{1}{2\pi} \oint_S \Omega_n(\boldsymbol{k}) \cdot d\boldsymbol{S}. \tag{2}$$

Here, $S$ is a two-dimensional (2D) surface encircling the Weyl point. $N_S$ is also known as the Chern number. For a conventional Weyl point, $N_S = \pm 1$, and its sign corresponds to the chirality of the Weyl points.

As discussed above, a 3D Weyl point occurs when either time-reversal or inversion symmetry is broken.[74] When both of them appear in solids, however, they could stabilize two Weyl points of opposite chirality at the same momentum, resulting in a fourfold degenerate point, denoted as a Dirac point (see Figure 1B). Materials that support such a degenerate point are called Dirac semimetals. The most prominent example of a Dirac semimetal is $A_3B$ (A = Na, K, Rb; B = As, Sb, Bi).[75] When considering up to the second order dispersion, the Dirac points can be described in the following form,

$$H_{\text{DSM}}(\boldsymbol{k}) = \epsilon(\boldsymbol{k}) + \begin{pmatrix} M(\boldsymbol{k}) & A(k_x + ik_y) & 0 & 0 \\ A(k_x - ik_y) & -M(\boldsymbol{k}) & 0 & 0 \\ 0 & 0 & M(\boldsymbol{k}) & -A(k_x - ik_y) \\ 0 & 0 & -A(k_x + ik_y) & -M(\boldsymbol{k}) \end{pmatrix}, \tag{3}$$

where, $M(k) = M_0 - M_1 k_z^2 - M_2(k_x^2 + k_y^2)$, $\epsilon(k)$ is the overall energy shift for the Dirac points. Being similar with the Weyl case, Dirac points can also be classified as type-I and type-II categories corresponding to the tilting degree of the Dirac points.[76,77] Here, the stability and position of Dirac points depend on $M(\boldsymbol{k})$. Generally constrained by crystallographic symmetries, there are still an additional seven types of Dirac points with different band dispersions and charge numbers.[78]

Besides the discrete points, the valence and conduction bands could cross along a one-dimensional curve (see Figure 1C). The curve may be in the shape of a closed loop, or take the form of a nodal line running across the Brillouin zone. The curve where the two bands cross is called a nodal line, and the materials that support such a nodal line in their low-energy region are called nodal line semimetals.[55] The essence of a 3D nodal line semimetal can be captured by a general effective model:

$$H_{NLSM}(\boldsymbol{k}) = \frac{1}{2m}(k_x^2 + k_y^2 - k_0^2)\sigma_z + v_z k_z \sigma_y, \tag{4}$$

where $m$ is the effective model along the radial direction, and $v_z$ is the Fermi velocity in the $k_z$-direction. Generally, nodal lines can be protected in at least four scenarios including chiral symmetry, combination of inversion and time-reversal symmetry, mirror symmetry, and glide mirror symmetry.

## 2.2 Topological surface states in topological semimetals

A striking manifestation of topological semimetals is the presence of nontrivial surface states. In Figure 1A, we provide a schematic diagram of two Weyl points of opposite topological charge in a 3D Brillouin zone. If a 2D slice is considered between these two Weyl points, the Chern number can be well defined ($|\mathcal{C}| = 1$). This is because any 2D slice between two oppositely charged Weyl points represents a 2D Chern insulator, which gives rise to a chiral edge state at its boundary. On the contrary, the Chern number is changed when encountering a Weyl point, say from 1 to 0, without edge states appearing at the boundary. Consequently, an open-edge state can be derived between these two Weyl points, denoted as the Fermi arc. According to the bulk-surface correspondence, the Fermi arc is a manifestation of Weyl semimetals.[79,80]

Two oppositely charged Weyl points are stabilized at the same point, resulting in a Dirac point. Given this picture in the bulk band structure, it is natural to expect that the hallmark of the Dirac semimetals is also the Fermi arc. Distinct from its Weyl counterparts, a Dirac point is characterized by double Fermi arcs,[81,82] as shown in Figure 1B. Since a Dirac point needs symmetry protection and carries a zero topological charge, its Fermi arcs are more subtle compared with that of the Weyl points. In comparison, a Dirac point that carries a nonzero topological charge has a topologically protected Fermi arc. The number of Fermi arcs is also equal to the absolute value of the topological charge, which may bring many more physical properties associated with Fermi arcs.

Weyl and Dirac semimetals have one-dimensional (1D) Fermi arc surface states. Nodal line semimetals also host nontrivial surface states, called drumhead surface state,[83-85] as shown in Figure 1C. In general, drumhead surface states appear within/out of the projection of the nodal lines, which depends on the boundary conditions. Nodal lines with higher-order dispersions also have corresponding surface states. The surface states in some cases are not protected, however, such as quadratic nodal lines, because of their trivial Berry phase. Compared with surface states for the topological semimetals with nodal points, this drumhead surface state is a 2D surface state, giving rise to a higher density of states on the surface.

### 3. Weyl topological catalysts

Weyl semimetals are the first category of topological semimetals for which there is evidence of catalytic enhancement for the HER. In Weyl semimetals, the Weyl fermions occur in pairs, and their number is determined by the specific symmetry of the material.[54] Weyl fermions features novel surface states with Fermi arcs. Note that, in 2017, Rajamathi *et al.* experimentally confirmed that a family of Weyl semimetals including NbP, TaP, NbAs, and TaAs are excellent HER catalysts.[56] They pointed out that the obtained high catalytic activity can be attributed to the presence of robust Fermi surface states and the occurrence of large carrier mobility at room temperature, both of which arose from the projections of the Weyl fermions in the bulk.

To verify their proposal, they examined the catalytic activity of these Weyl semimetals towards the dye-sensitized HER. Figure 2A shows the catalytic mechanism. They took the dye Eosin Y (EY) as the photon capture system. When the solar light is captured by EY, the resultant electrons would be excited and could transfer to the surface of the catalyst. Such excited electrons participate in the chemical transformation of $H^+$ to $H_2$ in water. In a Weyl semimetal, the surface Fermi arc arising from the bulk Weyl cones can serves as an active electron channel for the electron transfer. Figure 2B shows the position of these Weyl semimetals in a volcanic scheme. They found that NbP and TaP show good catalytic activity with small Gibb's free energies ($\Delta G_{H^*}$), which are comparable with Ni and Pt. Figure 2C specifically shows a comparison of the HER activity of all considered Weyl semimetals (NbP, TaP, NbAs, and TaAs), as obtained from single crystal powders supported by an intermediate dye addition. They found that the four Weyl topological catalysts all show high HER activity. Among them, NbP can be viewed as the best topological catalyst, where the $H_2$ value produced per gram was as high as 3520 μMol/g, combined with a quick reaction after adding the powdered single crystal dye. They argued that the HER catalytic enhancement in Weyl semimetals is highly responsible to the Fermi arcs on the surfaces. Unfortunately, such catalytic enhancement in these materials only acts in a limited level, because the Fermi arcs in traditional Weyl semimetals are generally short. On considering this, exploring new Weyl semimetals which carry long Fermi arcs may be an effective way to significantly improve the catalytic activity. To be noted, the Weyl semimetals in this work are focused on a photocatalystic process for HER, while other works which will be introduced in the following are for electrocatalytic process.

In 2022, Liu *et al.* proposed a novel Weyl semimetal category, namely, hybrid Weyl semimetals, which can naturally show long Fermi arcs and high catalytic performance towards the HER process.[57] In a hybrid Weyl semimetal, the band structure must host both type-I and type-II Weyl fermions in a single material.[70,71] Liu *et al.* focused their study on Ni-based materials. They first conducted a thorough materials screening of the Materials Project database combined with first-principles band structure calculations, and they found that the binary material NiSi is a good candidate as a topological catalyst with hybrid Weyl states. As shown in Figure 2D, they found that, near the Fermi level, NiSi possesses two Weyl points with diverse band slopes. Point $W_1$ has a traditional (namely type-I) Weyl cone, while $W_2$ has a tilted (namely type-II) Weyl cone, as shown in Figure 2E. Besed on first principles and an effective model analysis, they well demontrated that NiSi is a hybrid Weyl semimetal with both type-I and type-II Weyl cones near the Fermi level.

They further investigated the topological surface states for NiSi. Figure 2F shows the constant energy slices corresponding to the (010) surface at the Fermi level. They observed long Fermi arcs that connected the projected hybrid Weyl points. These Fermi arcs crossed throughout the Brillouin zone and were significantly longer than those in traditional Weyl semimetals. They found that the inclusion of spin-orbit coupling (SOC) doubles the number of Weyl points. Correspondingly, the Fermi arcs on the surfaces are also doubled (see Figure 2G). Furthermore, Liu *et al.* demonstrated that the long Fermi arcs make the surface of NiSi highly active for electron transfer in catalyzing the HER process. Figure 2H displays the likely catalytic mechanism in the hybrid Weyl semimetal NiSi. During these investigations, they found that the existence of Fermi arcs is robust against H adsorption (see Figure 2I). This suggests that the catalytic process has good stability. Most importantly, they found that the $\Delta G_{\mathrm{H}^*}$ for the HER on this surface is only 0.077 eV, significantly lower than the Weyl semimetals in the NbP category, as shown in Figure 2J. They also provided sufficient evidence on the relationship between the Fermi arcs and the catalytic activity by shifting the Weyl points under strain and electron doping.

The above-mentioned two works have demonstrated the feasibility of developing topological catalysts based on Weyl semimetals. These works also provide both theoretical and experimental evidence that the catalytic enhancement arises from the nontrivial Fermi-arc surface states. Based on these findings, exploring topological catalysts with a high intensity of nontrivial surface states is believed to be a route to realize the best catalytic enhancement.

## 4. Dirac and other nodal-point topological catalysis

### 4.1 Topological Dirac semimetals as HER catalysts

Beside Weyl semimetals, other nodal point semimetals such as Dirac semimetals have also been proposed as potential topological catalysts. Both Weyl and Dirac semimetals feature Fermi arc surface states. The number of Fermi arcs in Diracl semimetals is double comparing with Weyl counterparts, which may further facilitate the catalyzing process in Dirac semimetal. Recently, Yang *et al.* conducted a pioneering exploration of Dirac semimetals for catalyzing the HER process in $Nb_2S_2C$.[60] Their study involved skillful defect engineering in $Nb_2S_2C$, during which, they found that the coordination effect from both nontrivial surface states and ordinary surface states could maximally activate this material's electrocatalytic HER capability. As shown in Figure 3A, they compared the surface properties of the $Nb_2S_2C$ between the states before and after hydrogen adsorption. The cases of both pristine and defect-rich $Nb_2S_2C$ were taken into account. Before H adsorption, the Dirac points and the associated surface states occurred at about -0.2 eV. These electronic states are mainly contributed by the S-$3p_x$ and S-$3p_y$ orbitals. These chemically active S-$3p$ orbitals can interact with those of the adsorbed H. As a result, more occupied orbital derivatives from S-$3p_x$ and S-$3p_y$ arise in lower Fermi levels. This induces charge accumulation on the H atom and the S-$3p_x$ and S-$3p_y$ orbitals but charge exhaustion on the S-$3p_z$ orbitals, which will enhance the HER process. The catalytic enhancement can be further improved by the construction of S vacancies in $Nb_2S_2C$ by shifting the positions of Dirac points, with a further optimized $\Delta G_{H^*}$. This work has paved the way to activating the basic catalytic activity by combining nontrivial and trivial surface states.

Kong *et al.* also studied the relationship between a Dirac semimetal and catalytic efficiency, based on a type-II Dirac material, $VAl_3$.[61] From symmetry analysis and band structure calculations, they found that, in pure $VAl_3$ the surface states are located higher than the Fermi level, which limits its surface catalytic performance. They further modified the material with Ni doping. The final mixing ratio that had the best catalytic activity was found to be $V_{0.75}Ni_{0.25}Al_3$. Figure 3B compares the (100) surface states of $VAl_3$ and $V_{0.75}Ni_{0.25}Al_3$. They found that $V_{0.75}Ni_{0.25}Al_3$ has a larger energy increase due to its Fermi arc that that in $VAl_3$, indicating that more electrons are transferred from the Fermi arc to adsorbed H atoms in $V_{0.75}Ni_{0.25}Al_3$. Correspondingly, the $\Delta G_{H^*}$ of $V_{0.75}Ni_{0.25}Al_3$ is 0.115 eV, lower than that of $VAl_3$ (0.292 eV). Moreover, they performed a series of catalytic measurements on $V_{0.75}Ni_{0.25}Al_3$. They found that the material shows both a low overpotential of 175 mV (at the current density of 10 mA $cm^{-2}$) and a low Tafel slope of 68 mV $dec^{-1}$. These electrochemical properties are both better than those of unalloyed $VAl_3$. This work fully evidenced the feasibility of optimizing catalytic ability by regulating the topological surface states in Dirac semimetals.

## 4.2 High Chern number topological catalysts and catalytic mechanism

Beside Weyl and Dirac semimetals, some chiral nodal point semimetals have also been proposed as high-performance topological catalysts. In 2020, Yang *et al.* studied the topological and electrochemical properties of PtAl and PtGa.[62] Both materials have large topological charges with the Chern number of 4, arising from the symmetry of chiral materials and the multifold degeneracy of nodal point. The high Chern number leads to more pieces of Fermi arcs than traditional Weyl semimetals. In addition, these nodal points are not located at the same high symmetry momenta, making the Fermi arcs extremely long within a large energy window, as shown in Figure 3C. As presented, the calculated $\Delta G_{H*}$ values for the HER process in PtAl and PtGa are both as low as 0.13 eV. They further found that the Fermi arcs on the surface mainly arise from the Pt-*d* orbitals. This perfectly combines the activation effects of the topological surface states and Pt element. Besides the theoretical demonstrations, they also grew single crystals of PtAl and PtGa. Their electrochemical measurements revealed that the overpotential at the current density of 10 mA cm$^{-2}$ was only 13.3 mV and 14.0 mV for PtAl and PtGa, respectively. Their corresponding Tafel slopes were only 16 and 20 mV dec$^{-1}$. These results suggest that PtAl and PtGa have even better catalytic performance than Pt/C catalyst, which are consistent with the trends from the calculated $\Delta G_{H*}$ values. Their work has also proved that the topological surface states of the *d*-orbital contribution can weaken the interaction between the adsorbate and the substrate, which will effectively optimize the kinetics of H desorption.

In 2022, Li *et al.* described a new method to identify the location of active centers in a series of inorganic topological crystalline materials with symmetry-protected metallic surface states.[63] In these materials, the obstructed Wannier charge centers are pinned at unoccupied Wyckoff positions. The surfaces with such positions will then show metallic obstructed surface states (OSSs). Their studies were performed based on several representative single crystals, including 2H-MoTe$_2$ and NiPS$_3$ with the topological semimetal 1T'-MoTe$_2$ as a comparison. As shown in the Figure 3D, they found that the OSSs only appear on the edge surface of 2H-MoTe$_2$. In the case of the synthesized 2H-MoTe$_2$ and 1T'-MoTe$_2$ bulk crystals, at the current density of 1 mA cm$^{-2}$ for the whole crystal, the overvoltages of the edge and base surfaces were 374 mV, 398 mV. and 480 mV, respectively. These results fully showed that the catalytic activity of this material mainly arises from the edge surface. They also performed similar experiments on 1T'-MoTe$_2$. Their HER measurements found that the charge transfer resistance for its base surface is far lower than for its edge surface (89.9 Ω versus 157 Ω), showing higher conductivity for the base surface. The novel methodology developed in their work has not only advanced our understanding the fundamental mechanism in topological catalysts, but also can accelerate the development of new topological catalysts for the HER.

With the rapidly increasing number of topological catalysts, establishing the fundamental theory for the working mechanism of topological catalysts has become particularly important. Recently, Xu *et al.* proposed a parameter, namely, the projected Berry phase (PBP), which can link the topological signature in topological catalysts to their catalytic activity.[64] The PBP is an intrinsic parameter of materials and can be obtained from their electronic structures. The PBP can significantly affect the topological behavior and electrical properties, including carrier density, velocity, and mobility in solid materials. To preserve the time-reversal symmetry, their study focused on several nonmagnetic HER catalysts. They assumed that the PBP can reflect the most dominant properties involving different wave bands and transverse currents. As shown in Figure 3E, they established the exchange current density and PBP for the non-magnetic catalysts. They indeed obtained an excellent linear relationship between these two parameters. Their work illustrated the role of the bulk band structure in promoting electrochemical activity and gave a fresh understanding of the mechanism of HER catalysts.

## 5. Nodal line topological catalysts

### 5.1 3D nodal line topological catalysts

Compared with the 1D Fermi arc surface states in nodal point semimetals, nodal line semimetals are believed to possess much higher nontrivial electron density on the surface because they have 2D drumhead surface states. Considering this fact, topological catalysts with nodal lines may have better surface catalytic enhancement than nodal point ones. In 2018, Li *et al.* investigated for the first time the potential of nodal line semimetals as high-performance topological catalysts for the HER.[59] Their study was carried out on a family of TiSi-type materials. They argued that the bottleneck of Weyl topological catalysts lies in their low carrier density at the Fermi energy. They assumed that 3D nodal-line semimetals that have nontrivial drumhead surface states will break the bottleneck, because near the Fermi level, they have abnormally high carrier density on their surfaces, as shown in Figure 4A. Then, they verified their hypothesis by estimating the topological and electrochemical properties of the TiSi-type materials from first-principles. They found that these materials exhibit Dirac nodal lines in the $k_y = 0$ plane. They further found that, on the (010) surface, the density of states for the top layer is indeed much higher than that of the bulk phase, arising from the drumhead surface states. These states can provide robust active catalyzing sites for the HER process, as shown in Figure 4B. They compared the catalytic performance towards the HER on different surfaces, and found that the (010) surface, which has the full projection of nodal line, has the lowest $\Delta G_{H^*}$, suggesting a direct relationship between the nodal line and the HER catalytic performance. In particular, they found that the $\Delta G_{H^*}$ for TiSi is almost zero, suggesting extremely high catalytic activity. Their results fully demonstrated that these 3D nodal line semimetals have promising for serving as highly efficient HER catalysts.

Besides the HER, some nodal line semimetals were also proposed to catalyze the oxygen evolution reaction (OER) process. In 2019, Li *et al.* found that the magnetic nodal line semimetal $Co_3Sn_2S_2$ can be used as a highly active topological catalyst for the OER.[58] To be noted, this material features magnetic Weyl points under the ground ferromagnetic state[86], while at the room temperature, it is a paramagnetic phase with the nodal-line band struture[58]. Since this review is mainly focused on the topological catalysts for the HER, we herein only will have room for a brief discussion of this material. Unlike the Dirac nodal line in TiSi-type materials, $Co_3Sn_2S_2$ has a Weyl type nodal line. They found that $Co_3Sn_2S_2$ has topologically protected surface states mainly contributed by Co atoms, which induces high electronic conductivity and inherently high catalytic activity. Such topological surface states were evidenced by their high-resolution transmission electron microscopy (HRTEM) measurements. They further tested the catalytic performance on its single crystals. As shown in Figure 4D, the overvoltage for the OER is only 300 mV at a current density of 10 mA cm$^{-2}$. They found that the overvoltage value could be further reduced to 270 mV after crushing the large single crystals into small particles.

In 2020, Li *et al.* carried out the first experimental study on nodal line semimetals as topological catalysts for HER.[65] Their study focused on Pt-based materials, because Pt has good chemical stability, ultra-high electronic conductivity, and a proper *d*-band center for HER catalysis. They first conducted a thorough survey on the $Pt_xM_y$ binary compounds and chose $PtSn_4$ for further study because that material is suitable for easy large-scale growth and possesses high conductivity. As shown in the insets of Figure 4E, they demonstrated that $PtSn_4$ is a Dirac nodal line semimetal that has nontrivial topological surface states. To verify the exact relationship between its topological nodal-line state and the HER catalytic performance, they synthesized high quality $PtSn_4$ single crystals. As shown in Figure 4E, the electrochemical measurements showed that the overpotential value at the current density of 10 mA cm$^{-2}$ could be as low as 37 mV. This value is lower than that of Pt foil (71 mV) and is comparable to that of Pt/C nanostructures (28 mV). They also found the PtSn$_4$ sample had very good electrocatalytic stability, as determined by long-term and high-current-density electrochemical tests. Their results suggested that the nodal line topological catalyst PtSn$_4$ can serve as a favorable alternative to Pt catalyst because it can significantly reduce the use of Pt but preserves higher catalytic activity than pure Pt.

## 5.2 2D nodal line topological catalysts

The above works were focused on 3D topological semimetals with nodal lines as HER catalysts, where the drumhead surface states play the dominant role in catalytic enhancement. Note that 2D catalysts often show more striking properties than their 3D counterparts, including more active sites, easier strain application, more favorable catalytic activation, etc. Inspired by the success of 3D nodal line topological catalysts, in 2019, Tang *et al.* reported excellent catalytic performance towards various electrochemical reactions on Cu$_2$Si nanoribbons.[66] Their choice of Cu$_2$Si was based on the following facts: (i) the 2D form of Cu$_2$Si has already been experimentally synthesized; and (ii) angle-resolved photoelectron spectroscopy has provided good evidence that Cu$_2$Si monolayer is a 2D nodal line semimetal.[89] Tang *et al.* calculated the electronic band structure of Cu$_2$Si monolayer and evidenced the presence of a nodal line in 2D Cu$_2$Si. They paid the most attention to studying the catalytic performance towards CO$_2$ electroreduction. As shown in Figure 4C, they found that, on the edge of Cu$_2$Si where the topological edge states occur, the $\Delta G_{H*}$ of Cu$_2$Si nanoribbons for reducing CO$_2$ to CH$_4$ can be as low as 0.24 eV. Their results suggest that the topological edge states can favor catalytic performance towards CO$_2$ electroreduction. In addition, they also investigated the feasibility of Cu$_2$Si nanoribbons for the HER. They found that the $\Delta G_{H*}$ of the side reaction for the HER has the low value of 0.36 eV.

Quite recently, Wang *et al.* developed a series of ideal 2D nodal line topological catalysts with the $\Delta G_{H*}$ for the HER close to zero.[67] Unlike Cu$_2$Si monolayer, which possesses a closed nodal line, Wang *et al.* focused on 2D topological catalysts with open nodal lines. As shown in the right panel of Figure 4F, they assumed that 2D open nodal lines can achieve abnormally high electron density on the edge, because the topological edge states from open nodal lines can be distributed throughout the entire edge. To verify their assumption, they conducted a detailed investigation on 2D Cu$_2$C$_2$N$_4$. Their band structure calculations demonstrated the presence of an open nodal line and the associated long edge states in 2D Cu$_2$C$_2$N$_4$. They also demonstrated that the long edge states make its edge highly active for H adsorption, and it participates in the charge transfer during the HER process (see Figure 4F). Most remarkably, they found that the $\Delta G_{H*}$ for the HER was only 0.10 eV, which is almost equivalent to that of the noble metal Pt (0.09 eV). Figure 4F shows that Cu$_2$C$_2$N$_4$ is nearly situated at the top of the volcanic curve, suggesting high catalytic activity due to the long topological edge states. In addition, they also made a thorough material screening of the 2DMatPedia database for 2D nodal line topological catalysts. They screened out a total of 10 potential 2D topological catalysts with open nodal lines, which all show low $\Delta G_{H*}$ for the HER. Their work has provided a new pathway to realize high-performance 2D topological catalysts and also paved a viable route to developing HER catalysts that do not contain noble metals.

## 6. Conclusion

In this review, we provide a systematic overview of the recent progress in developing high performance HER catalysts utilizing topological semimetals. We briefly introduce the basic theory for different topological semimetals and their nontrivial surface states as the background for this review. From the intrinsic characteristics of topological semimetals, their advantages as HER catalysts can be summarized as follows: (i) linear band crossings in the bulk provide ultrahigh mobility for the electron conduction, which favors charge transfer during the HER process; (ii) the presence of nontrivial surface states makes the surface highly active towards the adsorption and diffusion of hydrogen; and (iii) the nontrivial surface states are robust against backscattering and small amounts of defects or nonmagnetic dopants, which favor high catalytic stability. These advantages have been evidenced since the application of Weyl semimetals as HER catalysts in 2017. Since then, various topological semimetals, including Dirac semimetals, multiple-fold nodal point semimetals, and 3D and 2D nodal line semimetals have been identified as high-performance topological catalysts. The motivation, the investigation processes, the main results, and the insights for these topological catalysts are systematically summarized. In addition, a few works concerned with the fundamental mechanisms behind the catalytic enhancements from topological aspects of matter are also summarized, which includes the theory of projected Berry phase, the length of Fermi arcs, and the strength of the surface density of states. This review can help researchers to understand the background, the course of development, and the most recent progress on topological catalysts for the HER.

## Acknowledgements

This work is supported by the National Natural Science Foundation of China (No. 12274112). The work is also funded by the Science and Technology Project of the Hebei Education Department, the Natural Science Foundation of Hebei Province, the S&T Program of Hebei (No. A2021202012), the Overseas Scientists Sponsorship Program of Hebei Province (C20210330), and the State Key Laboratory of Reliability and Intelligence of Electrical Equipment of Hebei University of Technology (No. EERI_ (No. EERI_PI2020009).

**Conflict of Interest**

The authors declare no conflict of interest.

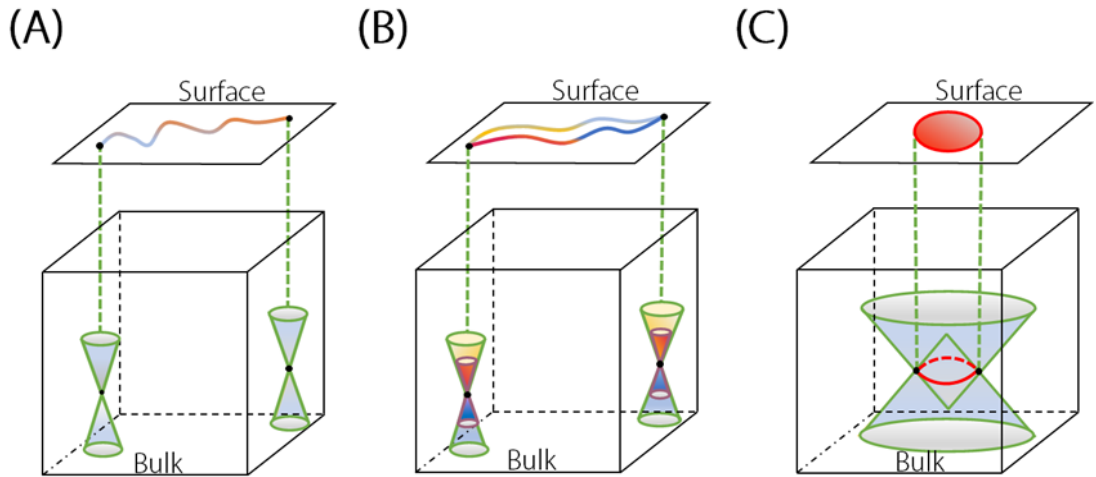

**Figure 1.** (A) Schematic diagram of a Weyl semimetal with one pair of Weyl points in the bulk and a Fermi arc on the surface. (B) is similar to (A) but represents a Dirac semimetal. Note: Weyl points have double band degeneracy, while Dirac points have fourfold band degeneracy. (C) is a schematic diagram of a nodal line semimetal with a nodal line in the bulk and drumhead-like states on the surface.

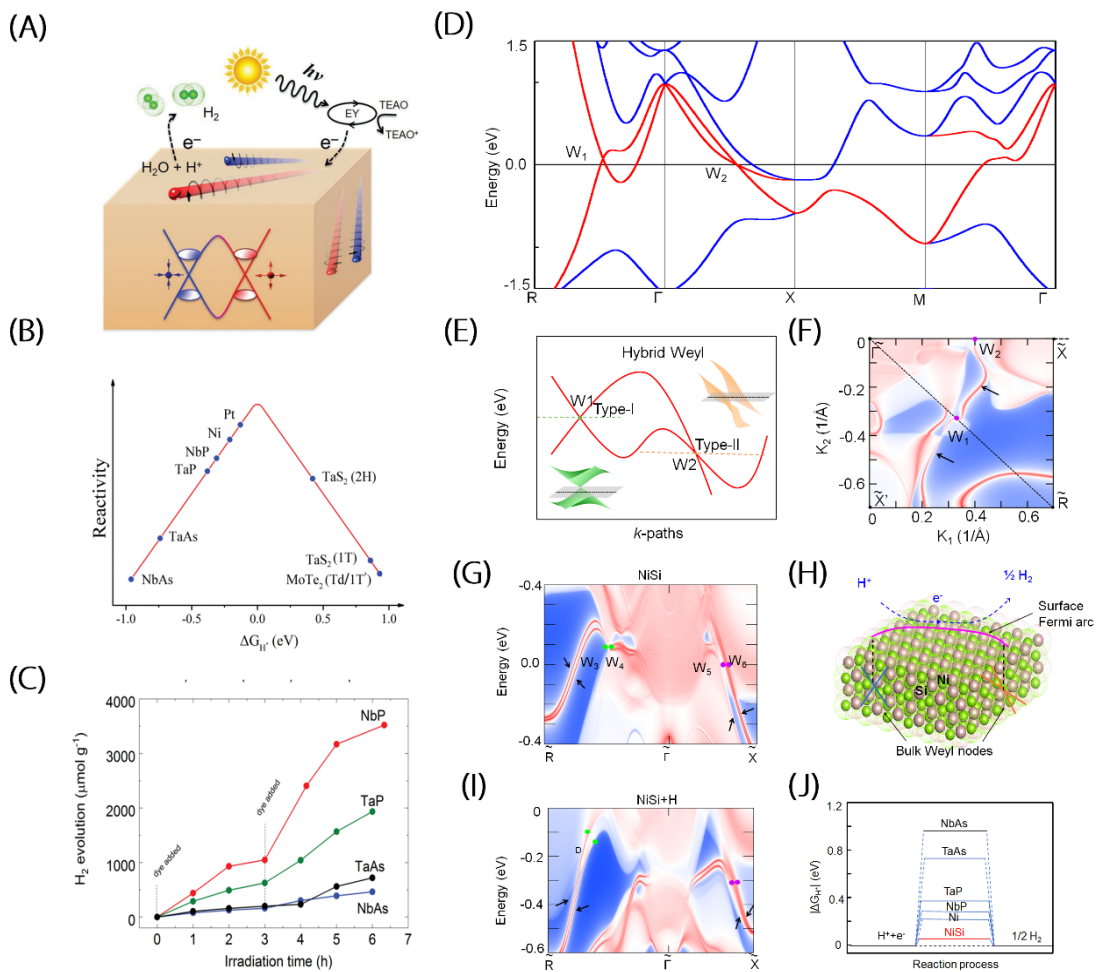

**Figure 2.** (A) Schematic diagram of a Weyl semimetal catalyzing dye-sensitized hydrogen evolution. (B) Relative activities of various HER catalysts following the volcanic scheme as a function of calculated $\Delta G_{H^*}$ on the surface of the catalyst. (C) Comparison of the hydrogen evolution activity of various Weyl semimetals with an intermediate dye addition. (A-C) Reprinted with permission.[56] Copyright 2017 Wiley-Blackwell. (D) Electronic band structure of NiSi. (E) Illustration of hybrid Weyl semimetal formed by combining type-I and type-II Weyl fermions. (F) Constant energy slices at the Fermi energy corresponding to the (001) surface. (G) Projected spectrum on the (001) surface of NiSi with SOC included. (H) Schematic illustration of the HER process on the NiSi surface. (I) The (001) surface states of NiSi after the hydrogen adsorption. (J) The free-energy diagram for some selected typical Weyl catalysts. (D)-(J) Reproduced from ref.[57] with permission from 2022 CELL.

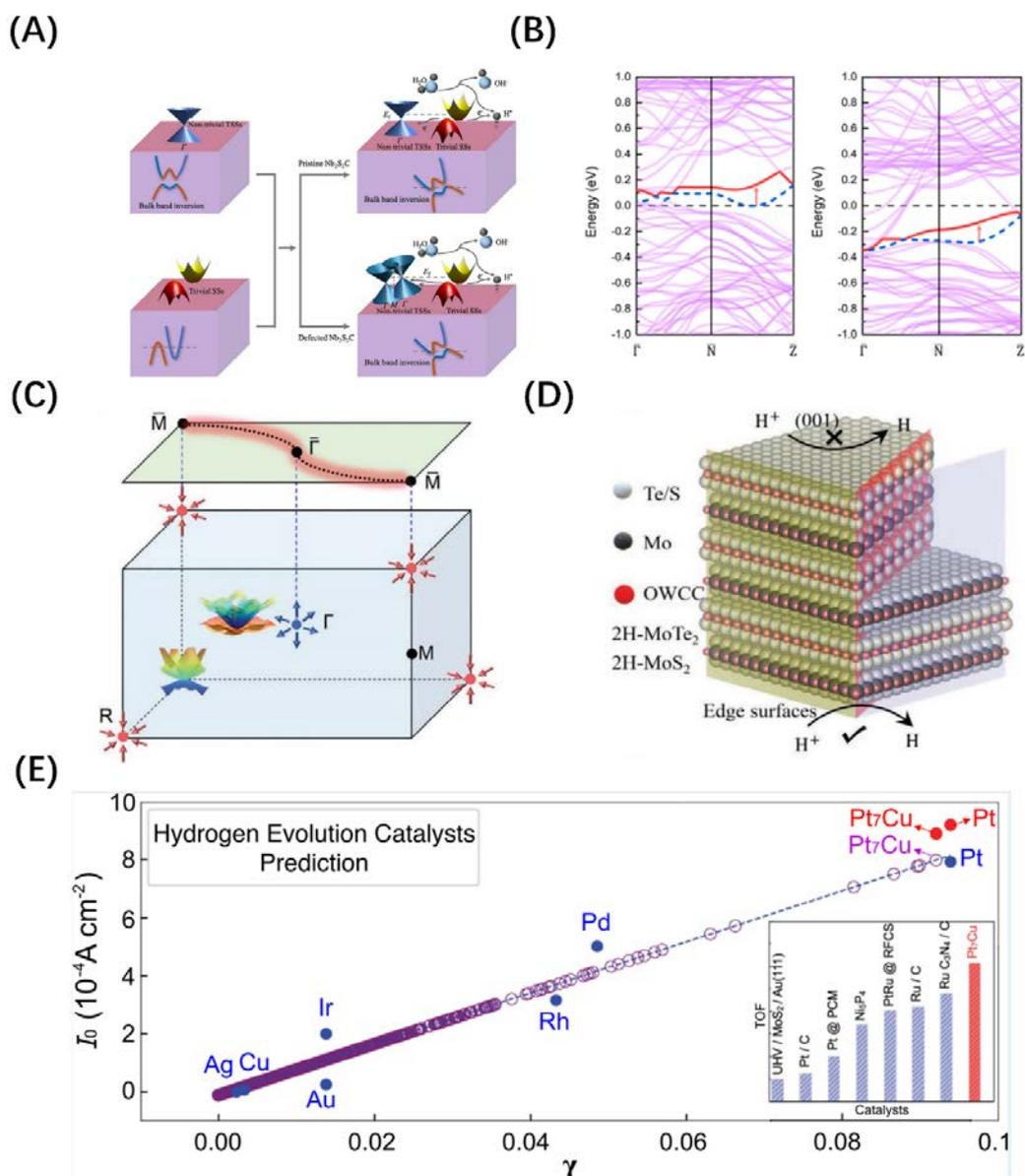

**Figure 3.** (A) The synergistic effect between non-trivial topological and trivial surface states for activating the electrocatalytic HER activity in Dirac semimetals. Reprinted with permission.[60] Copyright 2021 Elsevier. (B) Surface band dispersion before (the blue dashed line) and after (the red solid line) H adsorption onto the (100) surfaces for $VAl_3$ (the left panel) and $V_{0.75}Ni_{0.25}Al_3$ (the right panel). Reprinted with permission.[61] Copyright 2021 American Chemical Society. (C) Multifold nodal points and their surface states in PtAl. Reprinted with permission.[62] Copyright 2020 Wiley-Blackwell. (D) The crystal structure of 2H-MoTe2/2H-$MoS_2$ and the positions of obstructed surface states (OSSs). Reprinted with permission.[63] Copyright 2022 Wiley-Blackwell. (E) Theoretical results of the projected Berry phase (PBP) and exchange current density for different catalysts. Reprinted with permission.[64] Copyright 2020 American Chemical Society.

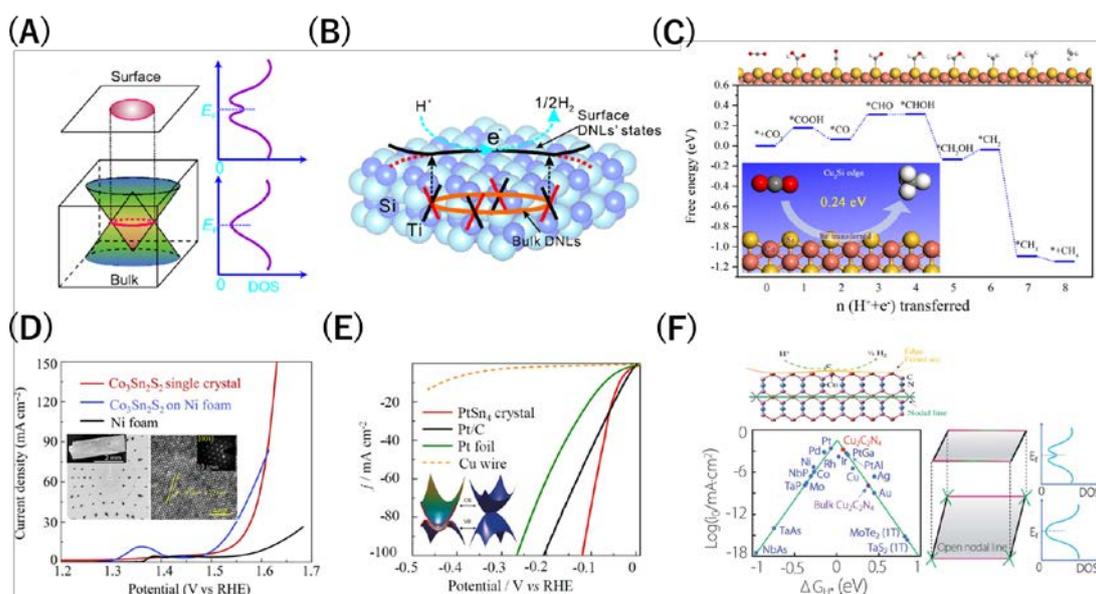

**Figure 4.** (A) The bulk and surface states of a nodal line semimetal. Its corresponding density of states is also provided. (B) Schematic illustration of bulk nodal-lines and surface states to provide an active plane and stable platform for the HER reaction in TiSi. (A) and (B) reprinted with permission.[59] Copyright 2017 Science China Press and Springer. (C) Free energy diagram of the 2D nodal-line semimetal $Cu_2Si$ for $CO_2$ reduction and the HER. The insets illustrate the catalytic mechanism. Reprinted with permission.[66] Copyright 2019 American Chemical Society. (D) The OER polarization curves for the topological nodal-line semimetal $Co_3Sn_2S_2$. The insets are the single-crystal XRD pattern and an HRTEM image. Reprinted with permission.[58] Copyright 2019 American Association for the Advancement of Science. (E) HER polarization curves for the Dirac nodal-line semimetal $PtSn_4$. The insets show the Dirac nodal line. Reprinted with permission.[65] Copyright 2019 Wiley-Blackwell. (F) The effect of open nodal lines and long edge states on the HER activity of 2D $Cu_2C_2N_4$ nanosheets, the volcanic curve, and the open nodal-line and the traversing Fermi arc with their corresponding DOS. Reprinted with permission.[67] Copyright 2021 The Royal Society of Chemistry.

**The table of contents entry**

The emerging research topic of utilizing topological semimetals as high-performance catalysts for the hydrogen evolution reaction has attracted rapid development in the last few years. This work provides a timely review of the background, motivation, investigation process, and fundamental mechanisms of topological catalysts, fully covering the Weyl, Dirac, multiple nodal point, and nodal line categories.

**Keyword**

hydrogen evolution reaction, electrocatalysis, topological semimetals, nontrivial surface states, topological catalysts

Lirong Wang[a,b,1], Ying Yang[c,1], Jianhua Wang[d,1], Wei Liu[a], Ying Liu[a,b], Jialin Gong[d], Guodong Liu[a,b], Xiaotian Wang[d,*], Zhenxiang Cheng[e,*], Xiaoming Zhang[a,b,*]

**Title**

Excellent catalytic performance towards the hydrogen evolution reaction in topological semimetals

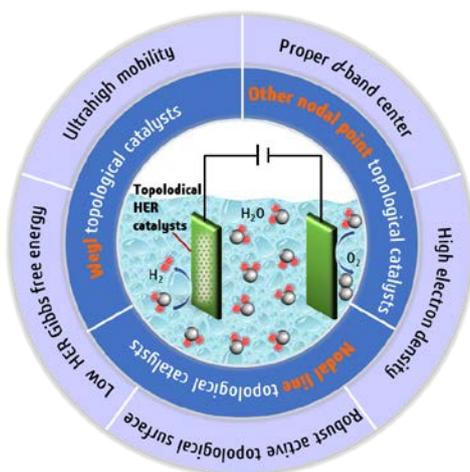

ToC figure (Please choose one size: 55 mm broad × 50 mm high **or** 110 mm broad × 20 mm high. Please do not use any other dimensions)